\documentclass[12pt]{article}

\usepackage{setspace}
\usepackage[margin=1in]{geometry}
\usepackage[T1]{fontenc}
\usepackage[utf8]{inputenc}
\usepackage{lmodern}
\usepackage{amsmath,amssymb,amsthm,mathtools,bm}
\usepackage{booktabs,array,multirow}
\usepackage{graphicx,adjustbox}
\usepackage{enumitem}
\usepackage{float}
\usepackage{needspace}
\usepackage[section]{placeins}
\usepackage{flafter}
\usepackage[font=small,skip=6pt]{caption}
\usepackage[round]{natbib}
\usepackage{xurl}
\usepackage{microtype}
\usepackage{tikz}
\usetikzlibrary{arrows.meta,positioning}
\usepackage[colorlinks=true,linkcolor=blue,citecolor=blue,urlcolor=blue]{hyperref}

\usepackage[normalem]{ulem}
\usepackage{xcolor}

\allowdisplaybreaks
\setdisplayskipstretch{1.0}
\floatstyle{ruled}
\newfloat{algorithm}{tbp}{loa}
\floatname{algorithm}{Algorithm}


\theoremstyle{plain}
\newtheorem{theorem}{Theorem}[section]

\newtheorem{proposition}{Proposition}[section]
\newtheorem{corollary}[theorem]{Corollary}
\theoremstyle{definition}
\newtheorem{assumption}{Assumption}[section]
\theoremstyle{remark}
\newtheorem{remark}{Remark}[section]

\newcommand{\E}{\mathbb{E}}
\renewcommand{\P}{\mathbf{P}}
\newcommand{\Var}{\operatorname{Var}}
\newcommand{\tr}{\operatorname{tr}}
\newcommand{\diag}{\operatorname{diag}}

\newcommand{\pto}{\xrightarrow{\P}}
\newcommand{\dto}{\xrightarrow{d}}
\newcommand{\cH}{\mathcal{H}}
\newcommand{\cN}{\mathcal{N}}
\newcommand{\op}{\operatorname{op}}

\title{Graph-Based Testing of Conditional Mean Independence with Applications to Large Language Model Evaluation}
\author{%
Mingshuo Liu\textsuperscript{1}\thanks{Liu and Zhu contributed equally.},\enspace
Qihua Zhu\textsuperscript{2}\footnotemark[1],\enspace
Jingchi Liao\textsuperscript{2},\enspace
Lu Tian\textsuperscript{1},\enspace
Doudou Zhou\textsuperscript{3}\thanks{Corresponding author: Doudou Zhou, ddzhou@nus.edu.sg.}\\[0.3em]
\footnotesize\textsuperscript{1}Department of Biomedical Data Science, Stanford University\\
\footnotesize\textsuperscript{2}Centre for Biomedical Data Science, Duke-NUS Medical School\\
\footnotesize\textsuperscript{3}Department of Statistics and Data Science, National University of Singapore}
\date{\vspace{-2em}}

\hypersetup{pdftitle={Graph-Based Testing of Conditional Mean Independence with Applications to Large Language Model Evaluation},pdfauthor={Mingshuo Liu, Qihua Zhu, Jingchi Liao, Lu Tian, Doudou Zhou}}
\makeatletter
\expandafter\gdef\csname r@app:application-details\endcsname{{G}{20}{}{}{}}
\expandafter\gdef\csname r@app:application-scale-null\endcsname{{A.1}{2}{}{}{}}
\expandafter\gdef\csname r@app:binary-edge-count-proof\endcsname{{J}{38}{}{}{}}
\expandafter\gdef\csname r@app:implementation-details\endcsname{{E}{18}{}{}{}}
\expandafter\gdef\csname r@app:k-sensitivity\endcsname{{C}{5}{}{}{}}
\expandafter\gdef\csname r@app:knn-computation\endcsname{{F}{20}{}{}{}}
\expandafter\gdef\csname r@app:non-euclidean-x-generation\endcsname{{B}{4}{}{}{}}
\expandafter\gdef\csname r@app:tabpfn-learner-sensitivity\endcsname{{C.3}{9}{}{}{}}
\expandafter\gdef\csname r@app:vector-null-generators\endcsname{{A}{1}{}{}{}}
\expandafter\gdef\csname r@app:zhang-comparison\endcsname{{D}{15}{}{}{}}
\expandafter\gdef\csname r@lem:power-knn-localization\endcsname{{S2}{37}{}{}{}}
\expandafter\gdef\csname r@tab:application-splits\endcsname{{S7}{24}{}{}{}}
\expandafter\gdef\csname r@tab:tabpfn-main-null\endcsname{{S3}{11}{}{}{}}
\makeatother
\begin{document}

\maketitle

\doublespacing

\begin{abstract}
We study whether a predictor $X$ provides additional information for
predicting a scalar response $Y$ after adjustment for other variables $Z$,
formalized by testing $\mathsf{H}_0:\E(Y\mid X,Z)=\E(Y\mid Z)$.
Existing methods often require estimating nuisance functions involving
$X$, which can be difficult for high-dimensional or structured predictors.
We propose a graph-based test that extends the edge count principle to
covariate adjustment. It represents $X$ through pairwise dissimilarities,
estimates only the baseline regression $\E(Y\mid Z)$, and aggregates
products of held-out residuals across graph edges. We develop sample-splitting
and two-fold cross-fitting versions. Under regularity conditions, both have
standard normal null limits, yielding analytic $p$-values, and are
consistent under graph-smooth and graph-rough alternatives. Simulations with
vector, image, and network predictors show good size control and power
advantages over competing procedures, with cross-fitting improving power
across the main alternatives. Applications examine human quality ratings,
mathematical error labels, and LLM response preferences. Our method detects
associations with quality ratings and an LLM judge's preferences after
adjustment for text length and other recorded characteristics.
\end{abstract}

\noindent\textit{Keywords:} variable significance; nearest neighbor graph; cross-fitting; non-Euclidean data; large language model evaluation

\section{Introduction}
\label{sec:intro}
Does the content of a language model response help predict the feedback it
receives, beyond its length and other recorded characteristics? This question
arises in both post-training and evaluation of large language models
(LLMs). Human feedback is used for instruction following and alignment with
human preferences \citep{ouyang2022training}, and for supervising mathematical
reasoning steps \citep{lightman2024verify}. LLM judges also provide automated
evaluations of model responses \citep{zheng2023judging}. Such feedback can be
biased, with consequences for both model training and evaluation. For example,
\citet{dubois2024length} document a systematic preference for longer responses
among LLM judges. \citet{chen2024odin} further show that, during reinforcement
learning, models can exploit a similar length bias in learned reward models
to obtain higher rewards by producing longer responses rather than improving
response quality. These findings motivate assessing the information in response content after accounting for length and other measured characteristics.

Let $Y$ denote a scalar feedback signal, and let $Z$ collect the
recorded characteristics used for adjustment. We use $X$ to denote a richer representation of the response,
such as a text embedding. Given independent and identically distributed
observations $(X_i,Y_i,Z_i)$, $i=1,\ldots,N$, the question can be formalized as
a test of conditional mean independence:
\begin{equation}
\label{eq:null}
\mathsf{H}_0:\quad \E(Y\mid X,Z)=\E(Y\mid Z).
\end{equation}
Rejection of \eqref{eq:null} provides
evidence that $X$ contains mean-predictive information after adjustment for
$Z$.

Testing \eqref{eq:null} is challenging in this setting. Whereas $Z$ contains
low-dimensional summaries of textual features, $X$ is typically a
high-dimensional learned embedding of response content. Accurately estimating
the augmented regression $m_1(x,z)=\E(Y\mid X=x,Z=z)$ or other
nuisance functions involving $X$ can therefore be difficult. This limits the direct applicability of many existing methods.
Classical kernel-based significance tests can require smoothing over the full
covariate vector, which becomes difficult when $X$ is high-dimensional
\citep{fan1996consistent,lavergne2000nonparametric}.
More recent procedures use flexible machine learning methods to accommodate
complex $X$, but their validity or power can still depend on accurate estimation of
nuisance functions involving $X$
\citep{cai2022model,dai2022significance,williamson2023general,cai2025test} or its conditional distribution \citep{zhang2025testing}.
The projected covariance method \citep{lundborg2024projected} instead relies on
learning a projection aligned with the additional conditional mean signal,
which can be challenging for high-dimensional embeddings.
Other residual-based tests smooth only over $Z$, but use bootstrap or
permutation procedures for inference
\citep{delgado2001significance,lavergne2015significance,bian2026partial}.

Beyond LLM evaluation, many other
modern applications call for testing conditional
mean independence with complex predictors. In neuroscience,
researchers may ask whether a brain connectivity network improves the
prediction of cognitive ability beyond standard demographic characteristics
\citep{Morris2022SCALARON}. In precision medicine, they may ask whether the
molecular structure of a drug, represented as a graph, affects expected
patient outcomes after adjustment for pretreatment characteristics
\citep{kaddour2021causal}. In both examples, $Z$ comprises relatively
low-dimensional demographic or clinical variables, whereas $X$ is a
non-Euclidean object. The complexity of $X$ in these settings motivates a different
way to capture the information in it.

Graph-based methods compare complex objects using similarity graphs, in which each observation is a vertex and edges connect pairs based on their similarity.
In two-sample testing, existing methods usually utilize an edge count principle which compares
within-sample and between-sample connectivity
\citep{friedman1979multivariate,schilling1986multivariate,rosenbaum2005exact,
chen2017new,chen2018weighted,zhou2023ranking, cai2024mates}. The same
connectivity principle also appears in change point detection, where
candidate temporal splits are assessed by comparing connectivity within and
across the resulting segments \citep{chu2019asymptotic,zhou2024changepoint,liu2026graphbmissing}.
For independence testing, \citet{liu2024generalized} combine pairwise
similarity and dissimilarity information, while \citet{zhu2025wise}
aggregate pairwise similarities with lag-dependent weights to test
serial independence.
\citet{chatterjee2024kernel} compare two conditional distributions using a
nearest neighbor graph on shared conditioning variables and a kernel on the
paired responses.
\citet{chatterjee2026conditional} use nearest neighbor graphs for
mean independence testing without covariate adjustment. They note
that extending their estimator to testing with additional covariates
generally requires further debiasing
\citep[Remark~4.3]{chatterjee2026conditional}.
We take a different approach to testing \eqref{eq:null}, using products
of held-out residuals along edges of a graph on $X$.

We propose CLEAR, which stands for \textbf{C}onditional mean independence
testing via \textbf{L}ocal \textbf{E}dge \textbf{A}ggregation of
\textbf{R}esidual products. CLEAR fits the baseline regression
$m(z)=\E(Y\mid Z=z)$ on $Z$ and sums negative products of residuals
over pairs of neighbors in $X$, defined by a $k$-nearest neighbor graph. This extends the edge count principle to
inference with covariate adjustment. For binary $Y$, the null has a conditional two-sample
interpretation \citep{lee2024frameworks}, detailed in
Section~\ref{subsec:binary-edge-count}. Without adjustment variables,
centering binary labels by their
pooled proportion gives a weighted edge count numerator, up to scale and an
additive constant.

CLEAR tests for systematic patterns in how residuals are distributed on the $X$-graph. Under the null, the product of residuals on each graph edge is centered at zero since removing
$m(Z)=\E(Y\mid Z)$ leaves oracle residuals with mean zero conditional on $(X, Z)$. In contrast, under an alternative, a smoothly
varying additional mean signal may give neighboring observations similar
residual means, leading to a positive edge product sum. If neighboring residual means instead have opposite
signs across sufficiently many edges, the edge product sum can be negative. We therefore use a two-sided test to detect
either pattern.

To separate baseline estimation from test evaluation, we develop a
single sample-splitting version (CLEAR-SS) and a two-fold cross-fitting
version (CLEAR-CF). Compared to single sample-splitting, the cross-fitted construction exchanges training and evaluation
roles and builds a
separate graph within each fold. This design allows every observation to contribute to the evaluation, potentially improving power while preserving the separation required for the validity analysis.

Our contributions are threefold. First, we develop a general framework
for testing conditional mean independence with complex predictors (Section~\ref{sec:setup}). The only fitted nuisance function is the baseline
regression $\E(Y\mid Z)$, while $X$ enters through pairwise
dissimilarities, accommodating Euclidean and non-Euclidean predictors.
Second, we establish standard Gaussian null limits for CLEAR-SS
and CLEAR-CF under graph regularity and prediction error conditions
(Section~\ref{sec:theory}). These results provide analytic $p$-values for both procedures.
We also establish consistency under graph-smooth and graph-rough
alternatives. Third, numerical experiments show good empirical size control
and substantial power gains for vector, image, and network predictors, compared with competing methods
(Section~\ref{sec:simulation}).
Section~\ref{sec:applications} examines human ratings, mathematical
reasoning feedback, and an LLM judge's preferences using text embeddings
\citep{kopf2023openassistant,lightman2024verify,chiang2024chatbot,zhou2026humananchored}.

\section{Method}
\label{sec:setup}

To separate the role of the graph from regression estimation, we first
construct the statistic with the baseline regression function $m$ known
(Section~\ref{subsec:graph-roughness}). We then estimate $m$ on an
independent training sample (Section~\ref{subsec:empirical-residual}) and
extend the construction to two-fold cross-fitting so that every observation
contributes to evaluation (Section~\ref{subsec:fw2-method}).
Finally, we discuss the application of CLEAR to conditional two-sample testing
(Section~\ref{subsec:binary-edge-count}) and the choice of baseline regression
and graph (Section~\ref{sec:implementation}).

Throughout this section, we use boldface for vectors in algebraic expressions and ordinary type for
their scalar components. Let $m(z)=\E(Y\mid Z=z)$, define the oracle
residuals $\varepsilon_i=Y_i-m(Z_i)$, and write
$\bm{\varepsilon}=(\varepsilon_1,\ldots,\varepsilon_n)^\top$.
We treat each fitted regression as a function of its training observations.
For square matrices $A_1,\ldots,A_q$, $\diag(A_1,\ldots,A_q)$ denotes
their block diagonal matrix. For a square matrix $A$, $\tr(A)$
denotes its trace.

\subsection{\texorpdfstring{$k$}{k}-nearest neighbor graph and oracle statistic}
\label{subsec:graph-roughness}

Consider $n$ i.i.d. observations $\{(X_i,Y_i,Z_i)\}_{i=1}^n$, and let
$\rho$ be a dissimilarity between values of $X$. We represent their
pairwise similarity by a symmetrized $k$-nearest neighbor graph, denoted by
$G_n=(V_n,E_n)$, where $V_n=\{1,\ldots,n\}$ and $E_n$ contains
unordered pairs of connected vertices. For each $n$, let $k=k_n$ be
an integer satisfying $1\le k\le n-1$, and suppress its dependence on
$n$. For each vertex $i$, select the $k$ observations $j\ne i$
with the smallest values of $\rho(X_i,X_j)$. We place an undirected edge
between $i$ and $j$ whenever either observation selects the other.
Distance ties are resolved by a fixed rule that does not depend on the
responses, as specified in Algorithm~\ref{alg:1}.

Let $A_n=(a_{ij})$ be the symmetric adjacency matrix, write $d_i$ for
the degree of vertex $i$, and let
$D_n=\diag(d_1,\ldots,d_n)$ be an $n\times n$ diagonal matrix and $L_n=D_n-A_n$.
The graph Laplacian quadratic form sums squared differences in oracle
residuals across edges:
$$
\bm{\varepsilon}^\top L_n\bm{\varepsilon}
=\sum_{\{i,j\}\in E_n}(\varepsilon_i-\varepsilon_j)^2.
$$
Under $\mathsf H_0$, $\E(Y_i\mid X_i,Z_i)=m(Z_i)$, so the oracle
residuals have conditional mean zero. Their conditional variances
$\sigma_i^2=\E(\varepsilon_i^2\mid X_i,Z_i)$ can still vary with the
covariates.  Let $\mathbf{X}=(X_i)_{i=1}^n$, $\mathbf{Z}=(Z_i)_{i=1}^n$ and $\mathbf{Y}=(Y_i)_{i=1}^n.$ Conditional independence across observations then gives
$$\E(\bm{\varepsilon}^\top L_n\bm{\varepsilon}\mid \mathbf{X}, \mathbf{Z})=\sum_{i\in V_n} d_i\sigma_i^2,$$ 
where the expectation is with respect to $\mathbf{Y}$ only. 
After ``centering'' $\bm{\varepsilon}^\top L_n\bm{\varepsilon}$ by subtracting the squared residual at each endpoint of every edge,  we therefore obtain 
\begin{equation*}
Q_n^\circ
=\bm{\varepsilon}^\top L_n\bm{\varepsilon}-\sum_{i \in V_n} d_i\varepsilon_i^2
=-2\sum_{\{i,j\}\in E_n}\varepsilon_i\varepsilon_j.
\end{equation*}
Given the covariates, every product
$\varepsilon_i\varepsilon_j$ with $i\neq j$ has expectation zero under
the null, and hence 
$$\E(Q_n^\circ\mid \mathbf{X}, \mathbf{Z})=0.$$

To see how a remaining mean signal affects the statistic, let
$g_i=\E(Y_i\mid X_i,Z_i)-m(Z_i)$ and
$u_i=Y_i-\E(Y_i\mid X_i,Z_i)$. Write
$\bm{g}=(g_i)_{i=1}^n$, $\bm{u}=(u_i)_{i=1}^n$, and
$\bm{\varepsilon}=\bm{g}+\bm{u}$. Since the $u_i$ are conditionally independent and centered at zero,
\begin{equation*}
\E(Q_n^\circ\mid \mathbf{X}, \mathbf{Z})
=-2\sum_{\{i,j\}\in E_n}g_i g_j
=\sum_{\{i,j\}\in E_n}(g_i-g_j)^2-\sum_{i \in V_n} d_i g_i^2.
\end{equation*}
Signals of the same sign at neighboring vertices shift the statistic
downward, whereas sufficiently strong opposite signs shift it upward.
We retain both directions in a two-sided~test.

Under the null, products from disjoint edges are conditionally independent. If two edges share vertex $i$, say $\{i,j\}$ and $\{i,\ell\}$, then
$\E(\varepsilon_i^2\varepsilon_j\varepsilon_\ell\mid \mathbf{X}, \mathbf{Z})=0$ because
$\varepsilon_i$, $\varepsilon_j$ and $\varepsilon_\ell$ are conditionally independent. Thus, distinct edge products are conditionally uncorrelated, giving
\begin{equation}
\label{eq:oracle-heterogeneous-variance}
v_n^2
=\Var(Q_n^\circ\mid \mathbf{X}, \mathbf{Z})
=2\sum_{i\ne j}a_{ij}^2\sigma_i^2\sigma_j^2
=4\sum_{\{i,j\}\in E_n}\sigma_i^2\sigma_j^2.
\end{equation}

\subsection{CLEAR-SS: sample splitting}
\label{subsec:empirical-residual}

In practice, the baseline regression is unknown, so the residuals used by the test must be estimated. Sample splitting separates this estimation from evaluation: we randomly divide the $N$
observation indices into a training fold $\mathcal I_1$, used to fit
$\widehat m$, and an evaluation fold $\mathcal I_2$ of size
$n=|\mathcal I_2|$, used to construct the residuals, graph, and statistic.
The split is independent of the data.

After ordering the evaluation observations by their original indices and
relabeling them as $1,\ldots,n$, define
$r_i=Y_i-\widehat m(Z_i)$ and $\bm{r}=(r_1,\ldots,r_n)^\top$.
Conditional on the training information, the evaluation observations remain
i.i.d. Replacing $\sigma_i^2$ with $r_i^2$ in \eqref{eq:oracle-heterogeneous-variance} yields the estimator
\begin{equation}
\label{eq:vhat-method}
\widehat v_n^2
=4\sum_{\{i,j\}\in E_n}r_i^2r_j^2.
\end{equation}
On $\{\widehat v_n^2>0\}$, let
$\widehat v_n=(\widehat v_n^2)^{1/2}$ and set
\begin{equation}
\label{eq:normal-stat}
T_n=\frac{-2\sum_{\{i,j\}\in E_n}r_i r_j}{\widehat v_n}.
\end{equation}
Let $\Phi$ denote the standard normal distribution function and $z_q$
its $q$th quantile. We report the Gaussian $p$-value
$p_n=2\Phi(-|T_n|)$ and reject at level $\alpha$ when
$|T_n|>z_{1-\alpha/2}$. If $\widehat v_n^2=0$, all edge products
vanish, so we set $T_n=0$ and $p_n=1$. Theorem~\ref{thm:sample-split}
shows that this event has probability tending to zero under its assumptions.

\begin{algorithm}[!htbp]
\caption{CLEAR-SS: residual graph test with sample splitting}
\label{alg:1}
\begin{minipage}{\linewidth}
\small\setstretch{1.1}
\textbf{Input:} Data $\{(X_i,Y_i,Z_i)\}_{i=1}^N$, level $\alpha$,
dissimilarity $\rho$, optional neighborhood size $k$, training proportion,
and regression method.\\
\textbf{Output:} Statistic, Gaussian $p$-value, and rejection decision.
\begin{enumerate}[leftmargin=1.7em,itemsep=.4em,topsep=.5em,parsep=0pt]
\item Independently of the data, split the indices into a training fold
$\mathcal I_1$ and an evaluation fold $\mathcal I_2$. The default
training proportion is half.
If $k$ is unspecified, set
$k=\min\{n-1,\lceil n^{1/4}\rceil\}$, where $n=|\mathcal I_2|$.
\item Fit and tune $\widehat m$ using only $\mathcal I_1$. Compute
$r_i=Y_i-\widehat m(Z_i)$ for $i\in\mathcal I_2$.
\item Build the symmetrized $k$-nearest neighbor graph on the evaluation
predictors. Join $i,j$ if either selects the other, resolve distance ties
by a fixed rule that does not depend on the responses, and count each unordered edge once.
\item Compute $Q=-2\sum_{E_n}r_i r_j$ and
$\widehat v_n^2=4\sum_{E_n}r_i^2r_j^2$. Set $T_n=Q/\widehat v_n$, or $T_n=0$ if
$\widehat v_n=0$, and $p_n=2\Phi(-|T_n|)$.
\item Reject if $p_n<\alpha$.
\end{enumerate}
\end{minipage}
\end{algorithm}

\subsection{CLEAR-CF: two-fold cross-fitting}
\label{subsec:fw2-method}

The observations used for training in sample splitting can also contribute
to evaluation if we exchange the roles of the two folds. We therefore use
two balanced folds $\mathcal I_1,\mathcal I_2$, of sizes $n_1,n_2$,
and evaluate each with a regression trained on the other. For $s=1,2$,
let $\widehat m^{(-s)}$ be fitted on the opposite fold
$\mathcal I_{3-s}$ and form
$r_i=Y_i-\widehat m^{(-s)}(Z_i)$ for $i\in\mathcal I_s$.
Within each fold, we construct a graph using the same dissimilarity and neighborhood size, yielding adjacency matrices $A_s$ and edge sets $E_s$.
We sum their contributions over edges within each fold:
\begin{align}
\widehat v_N^2
&=4\sum_{s=1}^2\sum_{\{i,j\}\in E_s}r_i^2r_j^2,
\label{eq:fw-vhat}\\
T_N
&=\frac{-2\sum_{s=1}^2\sum_{\{i,j\}\in E_s}r_i r_j}{\widehat v_N}.
\label{eq:fw-stat}
\end{align}
Here $\widehat v_N=(\widehat v_N^2)^{1/2}$.

Because each graph connects only observations within its own
fold, the statistic is a sum of two fold contributions.
With a known baseline regression, the fold contributions are independent
conditional on the covariates and split, which explains the addition of their
variance estimates in \eqref{eq:fw-vhat}. With estimated baselines, the resulting block diagonal structure
lets us control estimation error separately within each evaluation fold
(Section~\ref{subsec:fw2-theory}).
The test then uses the same two-sided Gaussian rule as
CLEAR-SS, with $p_N=2\Phi(-|T_N|)$.

\begin{algorithm}[!htbp]
\caption{CLEAR-CF: residual graph test with two folds}
\label{alg:fw2}
\begin{minipage}{\linewidth}
\small\setstretch{1.1}
\textbf{Input:} Data, level $\alpha$, dissimilarity $\rho$,
optional neighborhood size $k$, and regression method.\\
\textbf{Output:} Statistic, Gaussian $p$-value, and rejection decision.
\begin{enumerate}[leftmargin=1.7em,itemsep=.4em,topsep=.5em,parsep=0pt]
\item Independently of the data, divide the indices into two balanced
folds $\mathcal I_1,\mathcal I_2$.
If $k$ is unspecified, let $n_{\min}=\min(|\mathcal I_1|,|\mathcal I_2|)$
and set $k=\min\{n_{\min}-1,\lceil n_{\min}^{1/4}\rceil\}$.
\item For $s=1,2$, fit and tune $\widehat m^{(-s)}$ on
$\mathcal I_{3-s}$, then compute residuals on $\mathcal I_s$.
\item Build one graph in each fold, using a common
$1\le k\le\min(n_1-1,n_2-1)$ and the rules in Algorithm~\ref{alg:1}.
\item Sum the two edge product numerators and the two variance estimates
as in \eqref{eq:fw-vhat}--\eqref{eq:fw-stat}. Compute
$T_N$ and $p_N=2\Phi(-|T_N|)$ as in
Algorithm~\ref{alg:1}.
\item Reject if $p_N<\alpha$.
\end{enumerate}
\end{minipage}
\end{algorithm}

\subsection{Application to the conditional two-sample test}
\label{subsec:binary-edge-count}

For binary responses, \eqref{eq:null} is the conditional
two-sample null: the two response groups have the same conditional
distribution of $X$ given $Z$ \citep{lee2024frameworks}.
The baseline regression is the propensity
$\pi(z)=\P(Y=1\mid Z=z)$. A label-1 pair contributes
$\{1-\pi(Z_i)\}\{1-\pi(Z_j)\}$ to the oracle edge product sum.
Its contribution is small when the labels were already highly predictable
from the adjustment variables. The following identities connect this
construction to two-sample inference.

\begin{proposition}[Binary interpretation and weighted edge counts]
\label{prop:binary-edge-count}
Suppose $Y\in\{0,1\}$ and $0<\pi(Z)<1$ almost surely. Then
\eqref{eq:null} is equivalent to $Y\perp X\mid Z$, and hence to
$$
\mathcal L(X\mid Z,Y=1)=\mathcal L(X\mid Z,Y=0)
\quad\text{for almost every }Z.
$$
For any graph with edge set $E$ and any constant $\pi$, let
$R_{11}$ and $R_{00}$ be its two within-label edge counts. Then
\begin{equation}
\label{eq:binary-edge-count}
\sum_{\{i,j\}\in E}(Y_i-\pi)(Y_j-\pi)
=(1-\pi)R_{11}+\pi R_{00}-\pi(1-\pi)|E|.
\end{equation}
\end{proposition}

The proof appears in Supplementary Material~\ref*{app:binary-edge-count-proof}. Taking $\pi$ to be the pooled label proportion identifies the first two
terms in \eqref{eq:binary-edge-count} with the proportion-weighted
edge count statistic of \citet{chen2018weighted}. Their paper also studies
the finite-sample weights $(n_0-1)/(n_0+n_1-2)$ and
$(n_1-1)/(n_0+n_1-2)$, where $n_1,n_0$ are the group sizes.
CLEAR uses baseline predictions from the training fold and the variance
estimate in \eqref{eq:vhat-method} for its Gaussian test.

\subsection{Choice of regression and graph}
\label{sec:implementation}

Our framework accommodates a wide range of regression learners, provided
that the resulting baseline estimator satisfies the validity conditions in
Section~\ref{sec:theory}. Examples include generalized additive models
(GAMs), random forests, gradient boosting, Tabular Prior-data Fitted Network (TabPFN)
\citep{hollmann2025tabpfn}, and linear regression.
The main numerical study and applications use the same GAM for $\E(Y\mid Z)$. This permits nonlinear effects of the adjustment variables
while keeping the baseline learner fixed across settings. Supplementary
Material~\ref*{app:implementation-details} specifies the spline basis,
response families, and treatment of discrete adjustment variables.
In Supplementary Material~\ref*{app:tabpfn-learner-sensitivity}, we replace
the GAM with TabPFN in S1--S4 to assess how the baseline learner affects
Type I error control (Table~\ref*{tab:tabpfn-main-null}).

The dissimilarity should reflect the geometry of $X$ relevant to the
application. Euclidean distance is our default for vector predictors,
while other choices include Mahalanobis, Manhattan, and cosine distances,
as well as dissimilarities induced by kernels and distances between images and
networks. Once the dissimilarity is fixed, $k$ determines the scale at
which the test compares residual means. We use the sparse neighborhood rule
$k=\min\{n-1,\lceil n^{1/4}\rceil\}$ for sample splitting,
where $n$ is the evaluation sample size.
For two folds, we replace $n$ by $\min(n_1,n_2)$ so that both graphs
use the same $k$. This rule keeps the graph sparse while allowing $k$ to
grow. Increasing $k$ also makes the sufficient regression accuracy
condition in Proposition~\ref{prop:plugin-rate} more demanding.
Supplementary Material~\ref*{app:k-sensitivity} further examines a broader range of
graph scales. Algorithms~\ref{alg:1} and \ref{alg:fw2} give the
complete procedures. Once the graph and residuals are available, computing
the statistic and its variance estimate requires $O(nk)$ operations for
$n$ evaluation observations. Supplementary
Material~\ref*{app:knn-computation} discusses the cost of constructing the graph.

\section{Theoretical properties}
\label{sec:theory}

Section~\ref{subsec:ss-theory} establishes a Gaussian null approximation
for the sample-split statistic under conditions on graph regularity and
baseline regression accuracy. Section~\ref{subsec:fw2-theory} extends this
result to the two-fold statistic, accounting for the dependence induced
by exchanging training and evaluation samples.
Section~\ref{subsec:power-theory} studies the power of both constructions
under graph-smooth and graph-rough alternatives.

Throughout the section, for a vector $\bm{x}$, $\|\bm{x}\|_2$ denotes its Euclidean
norm.
For a matrix $M$, $\|M\|_{\op}=\sup_{\|\bm{x}\|_2=1}\|M\bm{x}\|_2$ and $\|M\|_F=\{\tr(M^\top M)\}^{1/2}$ denote its operator and Frobenius norms,
respectively.
We use $\pto$ and $\dto$ for convergence in probability and
distribution. The symbols $O_p(a_n)$ and $o_p(a_n)$
mean boundedness in probability and convergence to zero in probability
after division by the positive deterministic sequence $a_n$, respectively.

\subsection{Null approximation with sample splitting}
\label{subsec:ss-theory}

For the analysis, let $B_n=-A_n$ and write the statistic in
\eqref{eq:normal-stat} as
$$
T_n = \frac{\bm{r}^\top B_n \bm{r}}{\widehat v_n}.
$$
Write $\delta_i=m(Z_i)-\widehat m(Z_i)$, so that
$r_i=\varepsilon_i+\delta_i$, and let $\bm{\delta} = (\delta_i)_{i=1}^n$ collect these prediction
errors. Let $\mathcal A_N^{\mathrm{SS}}=(\mathcal I_1,\mathcal I_2)$
be the random split and set
$\cH_n=\sigma\!\left(\mathcal A_N^{\mathrm{SS}},
(X_j,Y_j,Z_j)_{j\in\mathcal I_1},
(X_i,Z_i)_{i\in\mathcal I_2}\right)$.
Here $\sigma(\cdot)$ denotes the $\sigma$-field generated by its
arguments.
Under $\mathsf H_0$, the oracle errors
$\{\varepsilon_i\}_{i=1}^n$ are independent conditional on $\cH_n$, and their conditional means and variances are
$$
\E(\varepsilon_i\mid\cH_n)=0,\qquad
\E(\varepsilon_i^2\mid\cH_n)=\sigma_i^2.
$$
Here $\sigma_i^2$ retains its meaning as a conditional variance
from Section~\ref{subsec:graph-roughness}.

To establish the null approximation theory, we impose the following assumptions. Throughout
this section, a ratio is defined as zero when its denominator is zero.
\begin{assumption}[Conditional fourth moments]
\label{ass:error}
For some constant $C<\infty$,
$\max_i\E(\varepsilon_i^4\mid\cH_n)\le C$ with probability tending
to one.
\end{assumption}

\begin{assumption}[Graph regularity]
\label{ass:graph}
As $n\to\infty$, $\P(v_n^2>0)\to1$ and
\begin{equation*}
\frac{\tr(A_n^4)}{v_n^4}\pto0.
\end{equation*}
\end{assumption}

\begin{assumption}[Small regression error on the graph]
\label{ass:regression}
Set $\Sigma_n=\diag(\sigma_1^2,\ldots,\sigma_n^2)$. The evaluation prediction errors satisfy
\begin{equation}
\label{eq:delta-main-cond}
\begin{split}
\frac{\bm{\delta}^\top A_n\Sigma_n A_n\bm{\delta}}{v_n^2}\pto0,
\quad
\frac{\bm{\delta}^\top B_n\bm{\delta}}{v_n}\pto0,
\quad
\frac{1}{v_n^2}\sum_{i\ne j}a_{ij}^2
\left(\sigma_j^2+\delta_j^2\right)\delta_i^2\pto0.
\end{split}
\end{equation}
\end{assumption}

\begin{theorem}[Gaussian null approximation with sample splitting]
\label{thm:sample-split}
Under $\mathsf H_0$, suppose Assumptions~\ref{ass:error}--\ref{ass:regression} hold. Let
$T_n$ be defined by \eqref{eq:normal-stat}, with variance estimate
\eqref{eq:vhat-method} and the zero variance convention in
Section~\ref{subsec:empirical-residual}. Then we have, as $n \to \infty$,
$$
\P(\widehat v_n^2>0)\to1,\qquad T_n\dto\cN(0,1).
$$
\end{theorem}

To see how the conditions lead to this result, expand the fitted numerator:
$$
\bm{r}^\top B_n \bm{r}=\bm{\varepsilon}^\top B_n\bm{\varepsilon}
+2\bm{\varepsilon}^\top B_n\bm{\delta}+\bm{\delta}^\top B_n\bm{\delta}.
$$

Conditioning on $\cH_n$ fixes the graph and $\bm{\delta}$, leaving
only the evaluation outcomes random. Under $\mathsf H_0$ the corresponding
errors are independent and centered, so the oracle term
$\bm{\varepsilon}^\top B_n\bm{\varepsilon}$ is a quadratic form in
independent variables. Its edge products are uncorrelated, although
products that share a vertex are not independent.
Assumption~\ref{ass:graph} limits this dependence, and together with the
moment bound in Assumption~\ref{ass:error}, which holds automatically for
binary $Y$ because $|\varepsilon_i|\le1$, it yields a Gaussian limit for
$\bm{\varepsilon}^\top B_n\bm{\varepsilon}/v_n$.

The other two terms in the expansion arise from fitting $m$, as
does the change in the variance estimate when $r_i$ replaces
$\varepsilon_i$. Assumption~\ref{ass:regression} requires each of these
effects to be small relative to the oracle standard deviation $v_n$. Its
first condition controls the conditional variance of the cross term
$2\bm{\varepsilon}^\top B_n\bm{\delta}$, its second the shift
$\bm{\delta}^\top B_n\bm{\delta}$, and its third the change in
$\widehat v_n^2$. Because these graph weighted conditions are not easy to
check directly, Propositions~\ref{prop:degree-sufficient}
and~\ref{prop:plugin-rate} give sufficient conditions in terms of the
maximum degree of the graph and the root mean squared prediction error of
$\widehat m$.

\begin{proposition}[A sufficient condition for graph regularity]
\label{prop:degree-sufficient}
Let $\Delta_n=\max_i d_i$ be the graph's maximum degree. Suppose
$\Delta_n=O_p(k)$, $k=o(n)$, and, for some $c>0$,
$\P(v_n^2\ge c nk)\to1$. Then
Assumption~\ref{ass:graph} holds.
\end{proposition}

\begin{proposition}[Sufficient prediction error rates]
\label{prop:plugin-rate}
Under Assumption~\ref{ass:error} and the conditions of
Proposition~\ref{prop:degree-sufficient}, suppose for a deterministic
sequence $a_n\to0$,
$$
\frac1n\sum_i\delta_i^2=O_p(a_n^2),\qquad
a_n=o\{(nk)^{-1/4}\}.
$$
Then Assumption~\ref{ass:regression} holds.
\end{proposition}

Both propositions use the maximum degree through the operator norm
bound $\|A_n\|_{\op}\le\Delta_n$. For the oracle statistic, this bound
gives $\tr(A_n^4)\le\|A_n\|_{\op}^2\|A_n\|_F^2\le2\Delta_n^2nk$, so the
ratio in Assumption~\ref{ass:graph} is $O_p(k/n)$ when $\Delta_n=O_p(k)$
and $v_n^2$ is of order $nk$. For the regression terms, it gives
$|\bm{\delta}^\top B_n\bm{\delta}|\le\Delta_n\|\bm{\delta}\|_2^2$, which
prevents prediction errors at a few vertices from being amplified through
many incident edges. Under the rate in
Proposition~\ref{prop:plugin-rate}, $\|\bm{\delta}\|_2^2=O_p(na_n^2)$, so
the shift is $O_p(nka_n^2)$. Requiring it to be of smaller order than
$v_n\asymp(nk)^{1/2}$ leads to $a_n=o\{(nk)^{-1/4}\}$, and this shift is
the term that determines the rate. Bounding the maximum degree is a
common way to rule out hubs in graph-based test statistics
\citep{chen2017new}, and for nearest neighbor graphs the bound holds under
standard conditions. For
symmetrized Euclidean nearest neighbor graphs under the
geometric and density conditions of \citet[Proposition~30]{vonluxburg2014hitting}
and with $k/\log n\to\infty$, the maximum degree is $O(k)$ with high
probability.
The variance lower bound holds, for example, if
$\min_i\sigma_i^2\ge\tau$ for some fixed $\tau>0$ with probability
tending to one. Since $|E_n|\ge nk/2$, this gives
$v_n^2\ge4\tau^2|E_n|\ge2\tau^2nk$.

\begin{remark}[A simpler regression rate]
\label{rem:plugin-rate}
Proposition~\ref{prop:plugin-rate} gives a simple sufficient condition on
the global root mean squared prediction error of the baseline regression on
$Z$. For fixed $k$, it requires $a_n=o(n^{-1/4})$. With the default
$k\asymp n^{1/4}$, it becomes $a_n=o(n^{-5/16})$. Increasing the graph
neighborhood size $k$ makes this sufficient condition more demanding.
\end{remark}

\subsection{Null approximation with two folds}
\label{subsec:fw2-theory}

We next extend the null approximation theory to the two-fold construction. Let $\mathcal A_N^{\mathrm{CF}}=(\mathcal I_1,\mathcal I_2)$ denote
the random split. For $s=1,2$, let $A_s$ be the adjacency matrix on
$\mathcal I_s$, set $B_s=-A_s$, and write
$A_N=\diag(A_1,A_2)$ and $B_N=\diag(B_1,B_2)$. The opposite fold
$\mathcal I_{3-s}$ trains the regression evaluated on $\mathcal I_s$.
Let $\delta_i^{(s)}=m(Z_i)-\widehat m^{(-s)}(Z_i)$, and order
$\bm{\delta}_N$ and $\bm{\varepsilon}$ by fold, so that
$\bm{r}_N=\bm{\varepsilon}+\bm{\delta}_N$. The statistic in \eqref{eq:fw-stat} and
its oracle variance are, respectively,
$$
T_N=\bm{r}_N^\top B_N\bm{r}_N/\widehat v_N,\qquad v_N^2=4\sum_{s=1}^2\sum_{\{i,j\}\in E_s}\sigma_i^2\sigma_j^2.
$$

Define $\mathcal G_N=\sigma\!\left(\mathcal A_N^{\mathrm{CF}},
(X_i,Z_i)_{i=1}^N\right)$ and
$\mathcal H_{N,s}=\mathcal G_N\vee
\sigma(Y_j:j\in\mathcal I_{3-s})$.
Here $\vee$ denotes the smallest $\sigma$-field containing both
$\sigma$-fields.
Under $\mathsf H_0$, the oracle
errors $\{\varepsilon_i\}_{i=1}^N$ are independent conditional on
$\mathcal G_N$, and their conditional means and variances are
$$
\E(\varepsilon_i\mid\mathcal G_N)=0,\qquad
\E(\varepsilon_i^2\mid\mathcal G_N)=\sigma_i^2.
$$
Additionally, adding the opposite fold's outcomes to $\mathcal G_N$ to form
$\mathcal H_{N,s}$ leaves the conditional distribution of the errors in fold
$s$ unchanged.

\begin{assumption}[Conditional fourth moments for two folds]
\label{ass:fw-error}
For some constant $C<\infty$,
$\max_i\E(\varepsilon_i^4\mid\mathcal G_N)\le C$
with probability tending to one.
\end{assumption}

\begin{assumption}[Graph regularity for two folds]
\label{ass:fw-graph}
As $N\to\infty$, $\P(v_N^2>0)\to1$ and
\begin{equation*}
\frac{\tr(A_1^4)+\tr(A_2^4)}{v_N^4}\pto0.
\end{equation*}
\end{assumption}

\begin{assumption}[Regression error for two folds]
\label{ass:fw-regression}
Set
$\Sigma_s=\diag(\sigma_i^2:i\in\mathcal I_s)$. The prediction errors within each evaluation fold satisfy
\begin{equation}
\label{eq:fw-delta-cond}
\begin{split}
&\frac{\sum_s(\bm{\delta}^{(s)})^\top A_s\Sigma_s A_s\bm{\delta}^{(s)}}{v_N^2}
\pto0,\qquad
\frac{\sum_s(\bm{\delta}^{(s)})^\top B_s\bm{\delta}^{(s)}}{v_N}\pto0,\\
&\frac{1}{v_N^2}\sum_s\sum_{\substack{i,j\in\mathcal I_s\\i\ne j}}
a_{s,ij}^2\left\{\sigma_j^2(\delta_i^{(s)})^2
+(\delta_i^{(s)})^2(\delta_j^{(s)})^2\right\}\pto0.
\end{split}
\end{equation}
\end{assumption}

\begin{theorem}[Gaussian null approximation with two folds]
\label{thm:fw2}
Under $\mathsf H_0$ and Assumptions~\ref{ass:fw-error}--\ref{ass:fw-regression}, the
variance estimate and statistic in \eqref{eq:fw-vhat}--\eqref{eq:fw-stat}
satisfy, as $N\to \infty$,
$$
\P(\widehat v_N^2>0)\to1,\qquad T_N\dto\cN(0,1).
$$
\end{theorem}

Two-fold cross-fitting links the fitted regressions through their training
outcomes. We use two levels of conditioning to separate the oracle analysis
from control of regression error. Conditional on $\mathcal G_N$, both graphs are fixed
and all oracle errors remain independent, so the block diagonal oracle
statistic admits the same normal approximation used for
Theorem~\ref{thm:sample-split}. Conditioning on
$\mathcal H_{N,s}$ fixes the fitted regression for fold $s$ but leaves that
fold's errors centered and independent, so the regression error terms are
controlled one fold at a time.

Proposition~\ref{prop:fw-sufficient} gives sufficient conditions for
Assumptions~\ref{ass:fw-graph} and \ref{ass:fw-regression} of the same form
as those in Propositions~\ref{prop:degree-sufficient} and
\ref{prop:plugin-rate} for sample splitting, with $N$ replacing $n$.

\begin{proposition}[Sufficient conditions for two folds]
\label{prop:fw-sufficient}
Let $\Delta_{N,s}$ denote the maximum degree of the graph on fold $s$.
Suppose both fold sizes are of order $N$, both graphs use the same $k$,
$\max_s\Delta_{N,s}=O_p(k)$, $k=o(N)$, and
$\P(v_N^2\ge cNk)\to1$ for some $c>0$. Then
Assumption~\ref{ass:fw-graph} holds. Under
Assumption~\ref{ass:fw-error}, if also
$$
\frac1N\|\bm{\delta}_N\|_2^2=O_p(a_N^2),\qquad
a_N=o\{(Nk)^{-1/4}\},
$$
then Assumption~\ref{ass:fw-regression} holds.
\end{proposition}

\subsection{Power analysis}
\label{subsec:power-theory}
In this section, we establish consistency of both tests under conditions
on (1) the strength of the conditional mean signal, (2) its alignment with the
graph, and (3) the accuracy of the baseline regression. We use unified
notation for the two constructions.
For sample splitting,
set $M=n$, $B_M=B_n=-A_n$, $\bm{\delta}_M=\bm{\delta}$, $T_M=T_n$,
$\widehat v_M=\widehat v_n$, and $\mathcal J_M=\mathcal H_n$.
For two folds, set $M=N$, $B_M=B_N$, $\bm{\delta}_M=\bm{\delta}_N$,
$T_M=T_N$, $\widehat v_M=\widehat v_N$, and
$\mathcal J_M=\mathcal G_N$.
Define $g_i=\E(Y_i\mid X_i,Z_i)-m(Z_i)$, arrange
$\bm{g}$ in the same order as the residuals, and let $\Delta_M$ be the
largest degree across the relevant graph blocks. All blocks use the same
$k$.
Write $u_i=Y_i-\E(Y_i\mid X_i,Z_i)$. The errors $u_i$
are independent and centered conditional on $\mathcal J_M$.
As in Section~\ref{subsec:fw2-theory}, conditioning further on
$\mathcal H_{N,s}$ in the two-fold case leaves the distribution of the
errors in fold $s$ unchanged.

We first state conditions on the error moments, signal strength, and
baseline regression.

\begin{assumption}[Conditional fourth moments under the alternative]
\label{ass:power-error}
For some constant $C<\infty$,
$\max_i\E(u_i^4\mid\mathcal J_M)\le C$ with probability tending to one.
\end{assumption}

\begin{assumption}[Signal strength]
\label{ass:power-signal}
The signal satisfies
\begin{equation*}
\|\bm{g}\|_2^2\sqrt{k/M}\pto\infty,
\qquad \frac1M\sum_i g_i^4=O_p(1).
\end{equation*}
\end{assumption}

\begin{assumption}[Negligible baseline regression error]
\label{ass:power-regression}
The baseline regression errors satisfy
\begin{equation*}
\frac{\|\bm{\delta}_M\|_2}{\|\bm{g}\|_2}\pto0,
\qquad \frac1M\sum_i\delta_{M,i}^4=O_p(1).
\end{equation*}
\end{assumption}

Beyond signal strength, power also depends on how the additional mean signal
aligns with the chosen graph. We quantify this alignment by the graph Rayleigh
quotient
\begin{equation*}
\mathcal R_M(\bm{g})=\frac{\bm{g}^\top B_M\bm{g}}{\|\bm{g}\|_2^2}
=\frac{\displaystyle\sum_s\sum_{\{i,j\}\in E_s}(g_i-g_j)^2
-\displaystyle\sum_s\sum_{i\in\mathcal I_s}d_i g_i^2}
{\|\bm{g}\|_2^2},
\end{equation*}
where only the evaluation fold is included for sample splitting, and both
folds are included for cross-fitting.  The first term in the numerator is the usual variation of a
signal on the graph \citep[Section~II-E]{shuman2013emerging}, whereas the
second is signal energy weighted by vertex degree. Smooth signals with little edgewise
variation give a negative quotient. A positive quotient requires variation
large enough to exceed this degree weighted benchmark. The quotient is
invariant to rescaling $\bm{g}$, so it describes the signal's graph pattern
separately from its overall strength.

The next two results combine the preceding assumptions with negative and
positive graph margins expressed through $\mathcal R_M(\bm{g})$ to establish
consistency under graph-smooth and graph-rough alternatives, respectively.

\begin{theorem}[Consistency under graph-smooth alternatives]
\label{thm:consistency}
Suppose $\Delta_M=O_p(k)$,
Assumptions~\ref{ass:power-error}--\ref{ass:power-regression} hold, and for
some $\kappa_1>0$,
\begin{equation}
\label{eq:smoothness-margin}
\P\{\mathcal R_M(\bm{g})\le-\kappa_1 k\}\to1
\end{equation}
as $M \to \infty$.
Then $\P(\widehat v_M^2>0)\to1$, and for every
fixed $\alpha\in(0,1)$,
$\P\{T_M<-z_{1-\alpha/2}\}\to1$.
\end{theorem}

\begin{corollary}[Consistency under graph-rough alternatives]
\label{cor:rough-signal-consistency}
Under the same conditions as Theorem~\ref{thm:consistency}, replace \eqref{eq:smoothness-margin} by
\begin{equation*}
\P\{\mathcal R_M(\bm{g})\ge\kappa_2 k\}\to1
\end{equation*}
for some $\kappa_2>0$. Then $\P(\widehat v_M^2>0)\to1$, and, for
every fixed $\alpha\in(0,1)$,
$\P\{T_M>z_{1-\alpha/2}\}\to1$ as $M \to \infty$.
\end{corollary}

For smooth alternatives, the negative graph margin, referred to below as
the smoothness margin, can be checked using the lengths of graph edges.
Let $\mathcal S_M=\{2\}$ for sample splitting, with $n_2=n$
and $E_2=E_n$, and let $\mathcal S_M=\{1,2\}$ for two folds. Define
\begin{equation*}
h_M^2=\frac1{Mk}\sum_{s\in\mathcal S_M}
\sum_{\{i,j\}\in E_s}\rho(X_i,X_j)^2.
\end{equation*}
Because the number of edges is of order $Mk$, $h_M$ measures the
typical distance between connected predictor values.

\begin{proposition}[A smoothness margin for signals Lipschitz in $X$]
\label{prop:power-smoothness-margin}
Suppose that, for a deterministic sequence $L_{g,M}\ge0$, with
probability tending to one,
\begin{equation*}
|g_i-g_j|\le L_{g,M}\rho(X_i,X_j)
\quad\text{for every edge in each evaluation graph}.
\end{equation*}
If $\P(\|\bm{g}\|_2>0)\to1$ and
$\sqrt M L_{g,M}h_M/\|\bm{g}\|_2\pto0$, then
\eqref{eq:smoothness-margin} holds for every $\kappa_1\in(0,1)$.
\end{proposition}

For a fixed signal shape, write
$g_i=b_M q(X_i)$, where $b_M\ne0$ sets the amplitude, $q$ is Lipschitz,
$\E\{q(X)\mid Z\}=0$, and
$M^{-1}\sum_iq(X_i)^2\pto q_0^2>0$. If $L_q$ is a Lipschitz
constant for $q$, take $L_{g,M}=|b_M|L_q$. The condition in
Proposition~\ref{prop:power-smoothness-margin} then becomes
$$
\frac{\sqrt M L_{g,M}h_M}{\|\bm{g}\|_2}
=\frac{L_qh_M}{\{M^{-1}\sum_iq(X_i)^2\}^{1/2}}\pto0.
$$
The amplitude cancels, and the denominator tends to $q_0>0$.
Thus, shrinking edge lengths, $h_M=o_p(1)$, verify the smoothness margin
for this signal shape.

\begin{remark}[Localization of nearest neighbor graphs]
\label{rem:power-knn-localization}
Localization means that the typical distance between observations connected
by an edge tends to zero. For i.i.d. predictors from a fixed distribution in
fixed Euclidean dimension, with graphs built using Euclidean distance and
$\E\|X\|_2^{2+\eta}<\infty$ for some
$\eta>0$, Lemma~\ref*{lem:power-knn-localization} gives $h_M=o_p(1)$
when each graph block has size $n_s\asymp M$ and $k/n_s\to0$.
The latter condition allows the number of neighbors to grow, provided its
fraction of the observations in each block tends to zero.
If also $\sqrt M L_{g,M}/\|\bm{g}\|_2=O_p(1)$, multiplying this factor
by $h_M=o_p(1)$ gives
$\sqrt M L_{g,M}h_M/\|\bm{g}\|_2\pto0$, which supplies the convergence
condition in Proposition~\ref{prop:power-smoothness-margin}.
\end{remark}

\section{Numerical experiments}
\label{sec:simulation}
We compare CLEAR-SS (Algorithm~\ref{alg:1}) and CLEAR-CF
(Algorithm~\ref{alg:fw2}) with competing methods under
the null and alternatives.
Both estimate the baseline regression $\E(Y \mid Z)$ with the common
GAM and use two-sided Gaussian critical values. We set $k=\min\{n-1,\lceil n^{1/4}\rceil\}$, where
$n=N/2$ is the size of each evaluation fold, giving $k=4$ and $k=5$ for
$N=500$ and $N=1000$, respectively. Supplementary
Material~\ref*{app:k-sensitivity} examines sensitivity to the neighborhood size.
Within each replication, CLEAR-SS and CLEAR-CF use the same simulated data, fold assignment, graph metric, and regression fits. The main designs comprise four vector null
settings (S1--S4) and three image or network settings (S5--S7). The vector
power study uses the S2 and S3 baseline functions, each with three signal
shapes.

The comparison includes three families of competing procedures. VIM uses an $R^2$-based target that compares oracle predictive
performance with and without $X$ \citep{williamson2023general}. PCM learns
an $X$-dependent projection of the remaining mean signal and tests its
held-out covariance with the residualized outcome
\citep{lundborg2024projected}. We also include its repeated-split version,
$\mathrm{PCM}_M$. pMIT
regresses the baseline residual on $(X,Z)$ and tests whether its conditional
mean is zero. pMITe adds the power enhancement term of
\citet{cai2025test}, and pMITM and pMITeM repeat these procedures over multiple splits.
Implementation details are given
in Supplementary Material~\ref*{app:implementation-details}. All tests use
nominal level $5\%$, and each main study setting uses 500 Monte Carlo
replications. Supplementary Material~\ref*{app:zhang-comparison} gives two
comparisons with the generative neural network test (GNN) of
\citet{zhang2025testing}: a matched-sample-size null comparison under
S1--S4 and a power comparison under their linear and quadratic signal models.
Evaluating GNN over the full simulation grid is computationally expensive,
so we report these focused comparisons separately.

\subsection{Experiments with Euclidean predictors}
\label{subsec:euclidean-x}
\label{subsec:level}
\label{subsec:EucAlter}

We first assess Type I error control with vector-valued $X$, using Euclidean
distance to construct the graph. The sample sizes are $N \in\{500,1000\}$,
and the dimensions of $Z$ and $X$ are $d_Z=20$ and $d_X=30$, respectively.
In all null settings, the error term
$\omega\sim N(0,0.5^2)$ is independent of $(\bm{Z}^\top,\bm{X}^\top)^\top$.

The settings in Table~\ref{tab:null-settings-summary} vary the dependence
between $X$ and $Z$ and the form of $\E(Y \mid Z)$. S1--S3 are adapted
from examples in \citet{Yan2025MachineLearningAssistedCO}: they keep the
dependence between $X$ and $Z$ linear. Their baseline means are
linear in S1, with $m(z)=\sum_{j=1}^{5}z_j$, radial in S2, and
exponential in S3. S4 makes both the dependence and
the baseline mean nonlinear. The generators are given in Supplementary
Material~\ref*{app:vector-null-generators}.
Supplementary Material~\ref*{app:k-sensitivity} further examines Type I error
control under heavy-tailed and positively skewed adjustment variables.

\begin{table}[!htbp]
\centering
\renewcommand{\arraystretch}{1.2}
\caption{Summary of null settings for empirical size evaluation.}
\label{tab:null-settings-summary}
\begin{tabular}{ccc}
\toprule
Setting & Dependence between $X$ and $Z$ & Conditional mean of $Y$ given $Z$ \\
\midrule
S1 & Linear & Linear \\
S2 & Linear & Nonlinear radial \\
S3 & Linear & Nonlinear exponential \\
S4 & Nonlinear & Nonlinear \\
\bottomrule
\end{tabular}
\end{table}

CLEAR-SS and CLEAR-CF generally retain stable Type I error control in all four settings at both sample sizes
(Table~\ref{tab:vector-null-rejection}). PCM and pMIT also have rejection rates generally around 5\%, with the rates ranging from
$3.2\%$ to $9.0\%$, while $\mathrm{PCM}_M$ rarely rejects under any null. The remaining competitors exhibit varying degrees of overrejection in some nonlinear settings (S2--S4). 
The contrast is most pronounced under the exponential baseline S3: CLEAR-CF
rejects at $5.4\%$ and $6.4\%$ for $N=500$ and $1000$, compared with
$38.2\%$ and $29.2\%$ for pMITe and $95.6\%$ and $88.2\%$ for pMITeM.
VIM also overrejects in S3.

\begin{table}[!htbp]
\centering
\caption{Null rejection rates (\%) in S1--S4 at nominal level $5\%$. CLEAR-SS and CLEAR-CF use the common GAM, two-sided Gaussian critical values, and $k=4$ for $N=500$ and $k=5$ for $N=1000$.}
\label{tab:vector-null-rejection}
\begingroup
\small
\setlength{\tabcolsep}{3pt}
\renewcommand{\arraystretch}{1.08}
\begin{tabular*}{\textwidth}{@{\extracolsep{\fill}}l*{8}{r}@{}}
\toprule
 & \multicolumn{2}{c}{S1} & \multicolumn{2}{c}{S2} & \multicolumn{2}{c}{S3} & \multicolumn{2}{c}{S4} \\
\cmidrule(lr){2-3}\cmidrule(lr){4-5}\cmidrule(lr){6-7}\cmidrule(lr){8-9}
Method & $N=500$ & $N=1000$ & $N=500$ & $N=1000$ & $N=500$ & $N=1000$ & $N=500$ & $N=1000$ \\
\midrule
CLEAR-SS & 4.6 & 5.4 & 4.8 & 6.8 & 5.4 & 4.4 & 5.0 & 5.4 \\
CLEAR-CF & 6.2 & 6.8 & 5.8 & 6.4 & 5.4 & 6.4 & 7.8 & 5.4 \\
PCM & 7.8 & 6.2 & 6.2 & 9.0 & 6.8 & 5.4 & 3.8 & 5.0 \\
$\mathrm{PCM}_M$ & 0.0 & 0.2 & 0.0 & 0.2 & 0.0 & 0.0 & 0.0 & 0.0 \\
VIM & 2.8 & 2.0 & 2.0 & 2.0 & 15.4 & 12.2 & 0.4 & 2.2 \\
pMIT & 6.4 & 6.4 & 3.8 & 4.8 & 5.0 & 3.2 & 5.6 & 5.8 \\
pMITe & 8.4 & 10.6 & 5.6 & 6.8 & 38.2 & 29.2 & 7.8 & 7.0 \\
pMITM & 6.2 & 11.6 & 6.0 & 6.6 & 12.6 & 6.6 & 5.2 & 8.2 \\
pMITeM & 10.8 & 18.4 & 8.6 & 11.4 & 95.6 & 88.2 & 8.0 & 12.6 \\
\bottomrule
\end{tabular*}
\endgroup
\end{table}

To examine rejection under alternatives, we add a single-index signal in $X$, keeping the sample
sizes and dimensions of $(X,Z)$ from the null study. We generate
$(\bm{Z}^\top,\bm{X}^\top)^\top\sim \cN(0,\Sigma)$ with
$\Sigma_{jk}=0.3^{|j-k|}$. We set $t=\bm{X}^\top\bm{\beta}$, where the first 25
coordinates of $\bm{\beta}$ equal 0.5 and the remaining coordinates equal zero.
The error term $\omega\sim N(0,0.5^2)$ is independent of $(X,Z)$. We
write the response model as
$$
Y=b_0(Z)+H_a(t;c_a)+\omega,
$$
where $b_0$ is the baseline component and $H_a$ is the additional signal
carried by $X$.
We consider the radial baseline
$b_{0,\mathrm{rad}}(z)=(\sum_{j=1}^{5}z_j^2)^{1/2}$ and the exponential
baseline $b_{0,\mathrm{exp}}(z)=\sum_{j=1}^{5}\exp(z_j)$, reusing the
baseline functions from null settings S2 and S3, respectively. For each
baseline $b\in\{\mathrm{rad},\mathrm{exp}\}$, we denote
the three alternatives by $\mathrm{A}^{b}_1,\mathrm{A}^{b}_2,
\mathrm{A}^{b}_3$. Their signal shapes are
$$
\mathrm{A}^{b}_1: H_1(t;c_1)=c_1t,
\qquad
\mathrm{A}^{b}_2: H_2(t;c_2)=c_2t^2,
\qquad
\mathrm{A}^{b}_3: H_3(t;c_3)=\frac{c_3t^2}{1+|t|}.
$$

The linear and quadratic signals in $\mathrm{A}^{b}_1$ and
$\mathrm{A}^{b}_2$ are adapted from
\citet{zhang2025testing}. The ridge signal in $\mathrm{A}^{b}_3$ is
quadratic near zero and approximately linear in the tails. We use
$(c_1,c_2,c_3)=(0.06,0.02,0.18)$ for the radial baseline and
$(c_1,c_2,c_3)=(0.22,0.07,0.6)$ for the exponential baseline.

\begin{figure}[!htb]
\centering
\includegraphics[width=\textwidth,height=0.70\textheight,keepaspectratio]{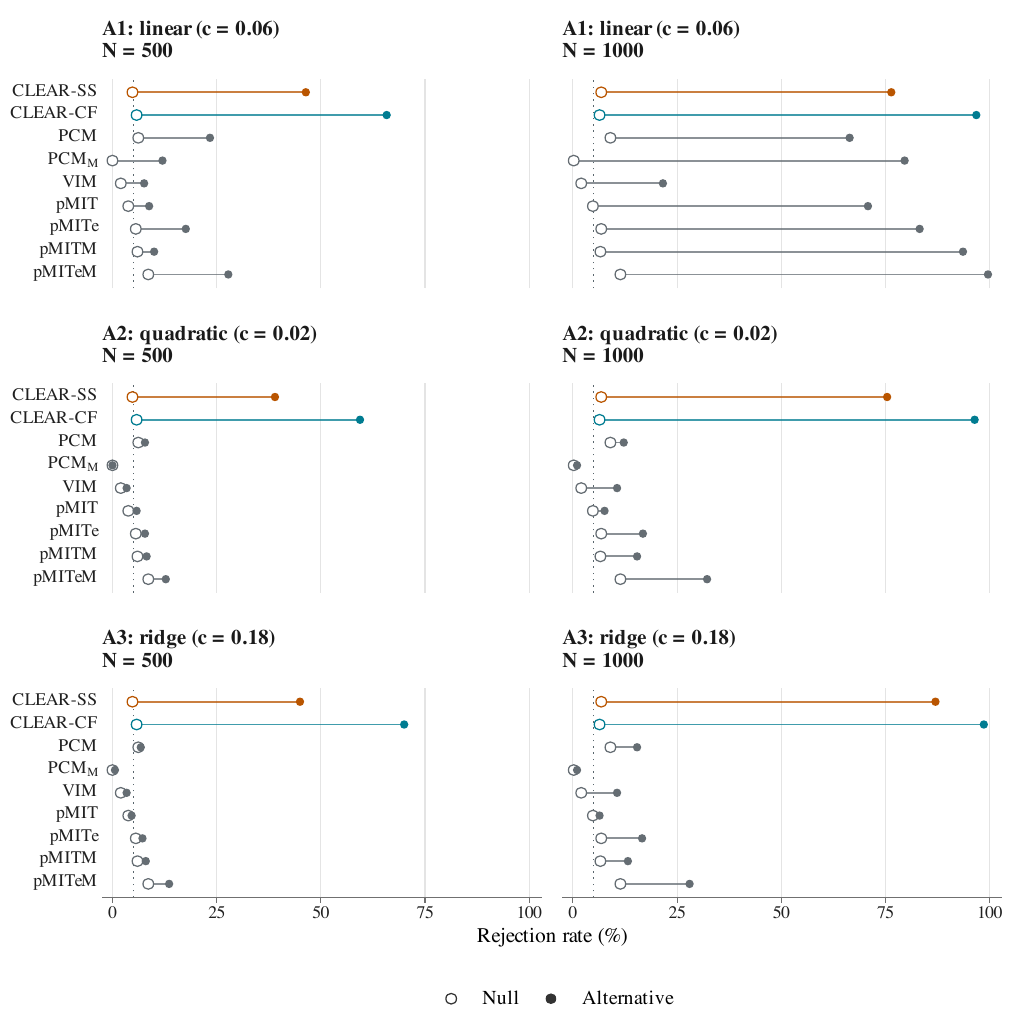}
\caption{Power comparisons under the radial baseline S2. Filled and open points show alternative and null rejection, respectively. Lines connect each pair. The dotted line marks $5\%$.}
\label{fig:matched-radial-rejection}
\end{figure}

Figure~\ref{fig:matched-radial-rejection} reports the results under
the radial baseline. CLEAR-CF has the highest power for all three signals
at $N=500$, and its advantage is largest for the two nonlinear signals.
At $N=1000$, it rejects $96.4\%$ of the quadratic and $98.6\%$ of the
ridge alternatives, compared with $32.2\%$ and $28.0\%$ for pMITeM, the
strongest competitor in these settings. For the linear signal, pMITeM achieves
a higher rejection rate of $99.6\%$, while its null rejection rate in S2 is $11.4\%$.
CLEAR-SS also exceeds every
competing method for these two signals at both sample sizes. The competing procedures detect the signal through fitted functions of \(X\). Quadratic and ridge functions of a 25-coordinate index may be harder to learn than a linear function, which could help explain the larger performance gap for the nonlinear signals.

Cross-fitting improves on CLEAR-SS for all three signals at both
sample sizes, by $11.6$ to $25.0$ percentage points. For the linear signal
at $N=500$, for example, power rises from $46.4\%$ to $65.8\%$.
Both versions use the same $k$, but
cross-fitting evaluates all $N$ observations rather than half of them.
For a fixed signal shape, $\|\bm g\|_2^2$ is proportional to the number
$M$ of evaluated observations, so the quantity $\|\bm g\|_2^2\sqrt{k/M}$
in Assumption~\ref{ass:power-signal}, which determines the order of the
standardized statistic under the alternative, increases by a factor of
about $\sqrt2$.

\begin{figure}[!htb]
\centering
\includegraphics[width=\textwidth,height=0.70\textheight,keepaspectratio]{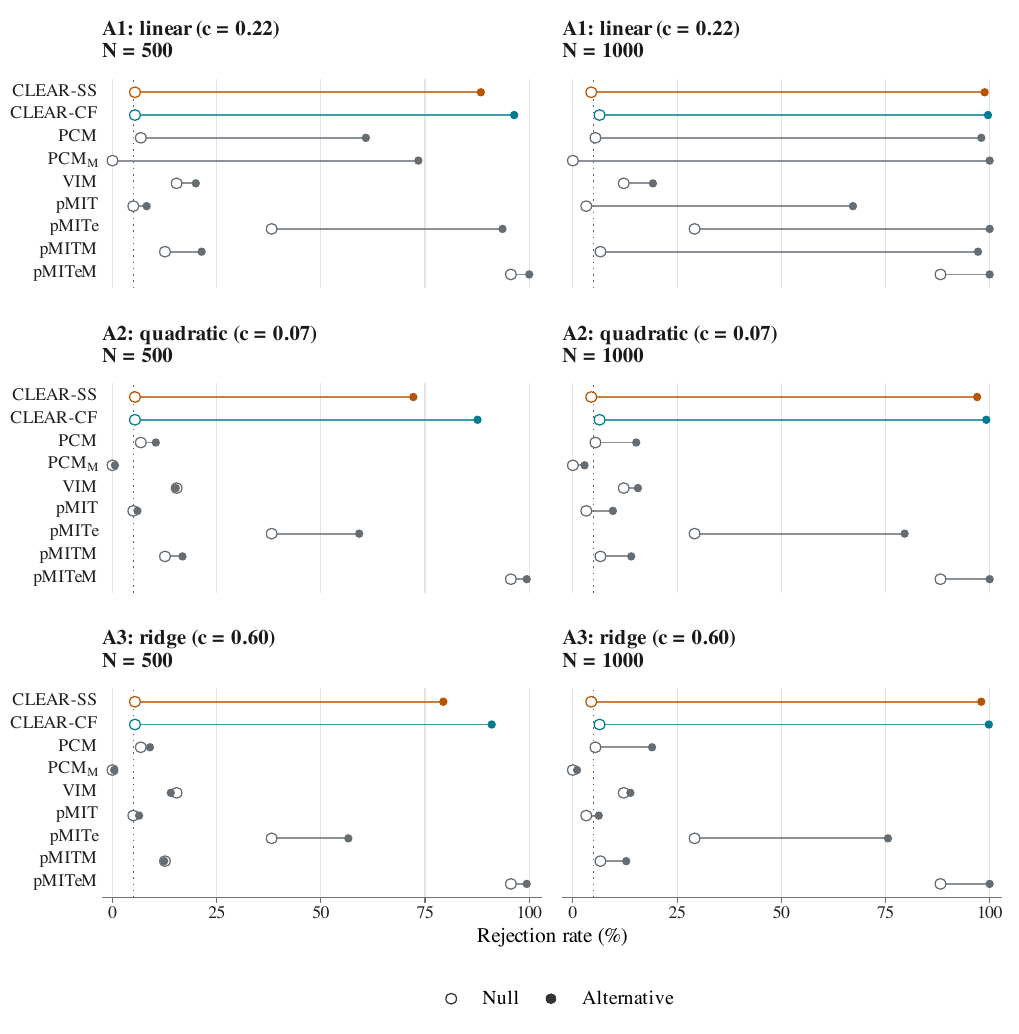}
\caption{Power comparisons under the exponential baseline S3. Symbols and the $5\%$ reference line follow Figure~\ref{fig:matched-radial-rejection}. The three alternatives share the corresponding S3 null.}
\label{fig:matched-exponential-rejection}
\end{figure}

Under the exponential baseline
(Figure~\ref{fig:matched-exponential-rejection}), the comparison must take
into account the null behavior of several competitors. In S3 at $N=500$,
VIM, pMITe, pMITM, and pMITeM reject $15.4\%$, $38.2\%$, $12.6\%$, and
$95.6\%$ of null replications (Table~\ref{tab:vector-null-rejection}), so
their alternative rejection rates overstate their power. Even so,
CLEAR-CF is more powerful than PCM, VIM, pMIT, and pMITM for all three
signals at both sample sizes, and more powerful than pMITe at $N=500$. As
under the radial baseline, the largest differences occur for the
nonlinear signals: at $N=500$, CLEAR-CF rejects $87.6\%$ of the quadratic
and $91.0\%$ of the ridge alternatives, compared with $10.4\%$ and $9.0\%$
for PCM. With the linear signal at $N=1000$, CLEAR-CF, $\mathrm{PCM}_M$,
and pMITe all reject more than $99\%$ of the alternatives.

At $N=1000$, CLEAR-CF rejects $99.2\%$--$99.8\%$ of the alternatives
for all three signals while its null rejection rate remains $6.4\%$.
Cross-fitting improves on sample splitting by $8.0$ to $15.4$ percentage
points at $N=500$. For $N=1000$, both CLEAR
versions are close to full power.

\subsection{Experiments with image and network predictors}
\label{subsec:non-euclidean-x}
We next study image- and network-valued $X$. Each observation is a triple
$(X_i,Z_i,Y_i)$, where $Z_i$ is a 20-dimensional adjustment vector and $Y_i$
is a scalar. In S5 and S6, $X_i$ is a $12\times12$ Gaussian image with
spatially correlated pixels. In S7, $X_i$ records the Gaussian weights of all
190 distinct edges of an undirected network whose 20 nodes are divided into
three communities. Edge weights in the same community pair block are correlated.
Each S7 network is one observation of the predictor. CLEAR separately constructs a
nearest neighbor graph across observations within each evaluation fold. In all
three settings, $X_i$ and $Z_i$ are independent, with
$Z_i\sim N(0,\Sigma_Z)$ and $(\Sigma_Z)_{ab}=0.3^{|a-b|}$. We use total
sample sizes $N\in\{500,1000\}$.

\begin{figure}[!htbp]
\centering
\includegraphics[width=0.9\textwidth]{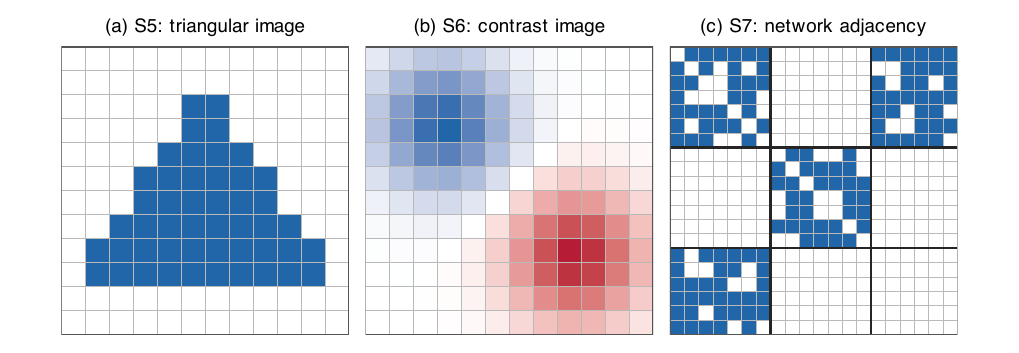}
\caption{Coefficient patterns used to construct the signals in S5--S7. S5
uses a positive central triangular region, S6 contrasts two localized regions
with opposite signs, and S7 assigns nonzero coefficients to selected edges in
two within-community blocks and one between-community block. Color intensity
shows coefficient magnitude, while blue and red show its sign. The S7 matrix is symmetric.
Diagonal entries are unused, and white entries have coefficient zero.}
\label{fig:non-euc-active-regions}
\end{figure}

The $X$-dependent signal $h_i$ is a linear score of the observed pixels or
edge weights. In S5, nonzero image coefficients form a central triangular
region. S6 contrasts two localized image regions with opposite signs. In S7,
the response coefficients are nonzero for selected edges in two
within-community blocks and one between-community block. The image patterns
are motivated by scalar-on-image regression \citep{Kang2016ScalaronImageRV},
and the network pattern by scalar-on-network regression
\citep{Morris2022SCALARON}. Figure~\ref{fig:non-euc-active-regions} shows
the three coefficient maps. Image scores are divided by their population
standard deviations, while the network coefficient vector has unit Euclidean
norm. Supplementary Material~\ref*{app:non-euclidean-x-generation} gives the
full generators and score normalization.

\begin{table}[!htbp]
\centering\small
\caption{Null rejection rates (\%) for GAM-based, two-sided CLEAR with image and network predictors. The nominal level is $5\%$. S5 and S6 share their null.}
\label{tab:gam-structured}
\begin{tabular}{llrrrr}
\toprule
 & & \multicolumn{2}{c}{$N=500$} & \multicolumn{2}{c}{$N=1000$} \\
\cmidrule(lr){3-4}\cmidrule(lr){5-6}
Setting & Distance & CLEAR-SS & CLEAR-CF & CLEAR-SS & CLEAR-CF \\
\midrule
S5 & $L^1$ & 4.6 & 4.4 & 5.6 & 6.8 \\
S5 & $L^2$ & 5.6 & 6.0 & 5.6 & 7.8 \\
S6 & $L^1$ & 4.6 & 4.4 & 5.6 & 6.8 \\
S6 & $L^2$ & 5.6 & 6.0 & 5.6 & 7.8 \\
S7 & $L^1$ & 5.6 & 5.4 & 6.2 & 6.8 \\
S7 & $L^2$ & 7.6 & 6.6 & 7.0 & 7.0 \\
\bottomrule
\end{tabular}
\end{table}

We generate the response as
$$
Y_i=m(Z_i)+\gamma_X h_i+\omega_i,
\qquad
m(Z_i)=\left(\sum_{j=1}^5 Z_{ij}^2\right)^{1/2},
\qquad
\omega_i\sim N(0,0.5^2),
$$
where $\omega_i$ is independent of $(X_i,Z_i)$. The radial baseline $m$ is
the one used in S2. Since $X_i$ and $Z_i$ are independent and $h_i$ is
centered, $\E(Y_i\mid Z_i)=m(Z_i)$ under both the null and alternatives.
Under the null, $\gamma_X=0$, so
$\E(Y_i\mid X_i,Z_i)=\E(Y_i\mid Z_i)=m(Z_i)$. Under the alternatives,
$\gamma_X=0.18$ for S5 and S6 and $\gamma_X=0.07$ for S7. S5 and S6 differ
only in their image coefficient maps, which do not enter the response when
$\gamma_X=0$. We therefore use the same null data for these two settings
within each replication.

Within each evaluation fold, CLEAR constructs the graph using entrywise
$L^1$ or $L^2$ distances between aligned pixels or network edge weights,
with $k=4$ at $N=500$ and $k=5$ at $N=1000$. The competing methods receive
the same 144 pixels or 190 edge weights as vector inputs.

\begin{figure}[!htb]
\centering
\includegraphics[width=\textwidth,height=0.70\textheight,keepaspectratio]{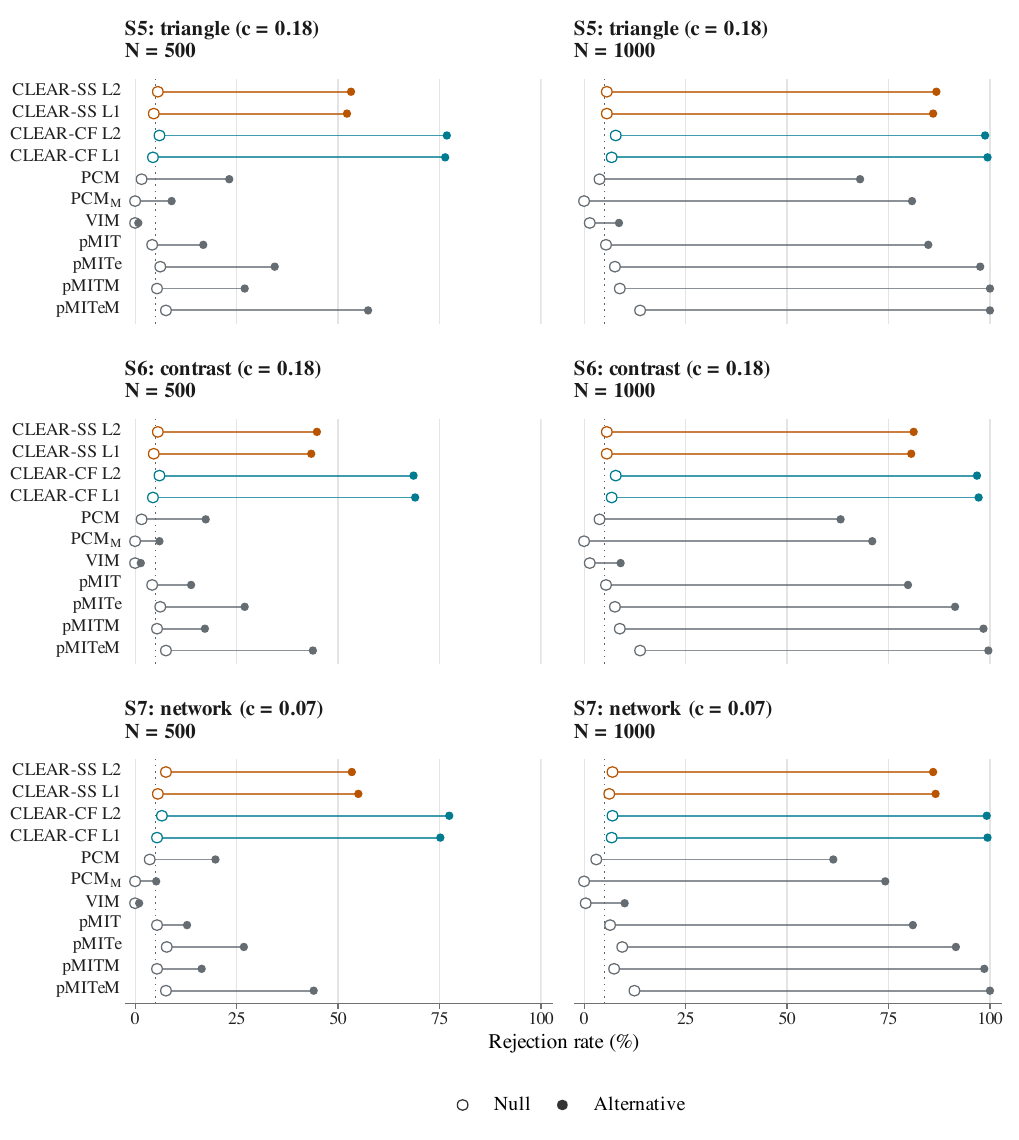}
\caption{Power comparisons for image and network predictors in S5--S7. CLEAR uses the indicated distance. Symbols follow Figure~\ref{fig:matched-radial-rejection}.}
\label{fig:matched-structured-rejection}
\end{figure}

At $N=500$, CLEAR-CF is the most powerful method in S5--S7 under
both distances (Figure~\ref{fig:matched-structured-rejection}). Its power
ranges from $68.6\%$ to $77.4\%$, which is $19.0$ to $33.4$ percentage
points above that of pMITeM, the strongest competitor in each setting,
and both CLEAR versions have null rejection rates near $5\%$ for images
and networks under both distances, although CLEAR-SS reaches $7.6\%$
in S7 under $L^2$ at $N=500$ (Table~\ref{tab:gam-structured}). The
advantage holds across the three signal patterns.
At $N=1000$, CLEAR-CF rejects $96.8\%$--$99.4\%$ of the
alternatives and exceeds PCM, $\mathrm{PCM}_M$, VIM, pMIT, and pMITe in
every structured setting. The repeated-split pMIT variants also approach
full power, but their null rejection rates are $7.4\%$--$13.8\%$,
whereas CLEAR-CF stays below $10\%$, at $6.8\%$--$7.8\%$. As in the vector settings, cross-fitting
improves on sample splitting, by $20.2$ to $25.6$ percentage points at
$N=500$ and by $12.0$ to $16.6$ at $N=1000$.
Supplementary Material~\ref*{app:k-sensitivity} examines sensitivity
to the neighborhood size and baseline learner.

\section{Applications to large language model evaluation}
\label{sec:applications}

We examine whether evaluations of language model responses vary with
text representations after adjustment for length and other recorded
characteristics. The first two applications concern human feedback:
quality scores for assistant replies and error labels for mathematical
solutions. The third asks the same question about an LLM judge's
preferences between responses. In each application, $Y$ records the
feedback, $X$ contains the text embeddings, and $Z$ contains the adjustment
variables, giving the conditional mean null in \eqref{eq:null}.

We encode complete texts with the Sentence-BERT model
\texttt{all-MiniLM-L6-v2} \citep{reimers2019sentence}\footnote{\url{https://huggingface.co/sentence-transformers/all-MiniLM-L6-v2}}
and construct Euclidean neighborhood graphs on the unit-normalized embeddings.
The components of $X$ are specified below. We compare CLEAR-SS and
CLEAR-CF with PCM and $\mathrm{PCM}_M$. We exclude VIM and the pMIT
variants because of their inflated rejection rates in the null
experiments in Supplementary Material~\ref*{app:application-scale-null}.
In the OpenAssistant and Arena null designs, keeping the observed $X$
and $Z$ but simulating $Y$ from a model that depends only on $Z$,
these methods reject in up to $81\%$ of
replications at the $5\%$ level, compared with $4.8\%$--$6.0\%$ for CLEAR.

For each application, we run every procedure $R=200$ times on the same
observations, varying the sample partitions. Table~\ref{tab:application-aggregate} reports
$p_{\mathrm{agg}}=\min\{1,2\,\operatorname{median}(p_1,\ldots,p_R)\}$
for each procedure
\citep{meinshausen2009pvalues}. If the $p$-values from each run are valid, this
rule controls Type I error even when the runs are dependent.
Supplementary Material~\ref*{app:application-details} gives the sample
construction, aggregation argument, and variation across runs.

\subsection{Human quality ratings in OpenAssistant}

OpenAssistant Conversations contains assistant replies written and annotated
by crowdworkers \citep{kopf2023openassistant}.\footnote{\url{https://huggingface.co/datasets/OpenAssistant/oasst1}}
We retain one accepted English reply to each distinct normalized English
initial prompt, with an available quality label. When several replies qualify,
a fixed rule that does not use quality scores selects one. All $N=3,400$
eligible prompts are used. Supplementary
Material~\ref*{app:application-details} gives the construction.

Here $Y$ is the released aggregated quality score, between zero and one.
We represent each reply by its 384-dimensional embedding, normalized to
unit length, and use this vector as $X$.
The five coordinates of $Z$ are prompt and response word and character
counts and the number of quality annotations, each transformed as
$\log(1+c)$. We test whether the conditional mean of the released quality
score varies with the response embedding beyond these summaries.

CLEAR-CF detects variation in mean quality scores over the response
embeddings beyond text length and annotation count, with an aggregated
$p$-value of $0.0020$.
CLEAR-SS also rejects, with $p_{\mathrm{agg}}=0.0356$, whereas PCM and
$\mathrm{PCM}_M$ give $0.0771$ and $0.0934$, respectively
(Table~\ref{tab:application-aggregate}).
CLEAR-CF yields $p<0.05$ in 196 of the 200 runs, compared with
111 for each PCM procedure
(Table~\ref*{tab:application-splits}).

\subsection{Human error labels in PRM800K}

We next examine error labels in PRM800K, which records human assessments
of model-generated mathematical solutions, including judgments on individual
reasoning steps
\citep{lightman2024verify}.\footnote{\url{https://github.com/openai/prm800k}}
From the released annotations, we retain one original solution for each
of $N=1,000$ distinct problems. The binary outcome $Y$ equals one when an
annotator identifies an error in that solution. It equals zero only when
all original steps were reviewed without a negative rating. Reviews that
end without a recorded error but leave steps unreviewed are excluded.

The predictor $X$ concatenates the embeddings of the problem and its
original solution into a normalized pair representation. The 11 coordinates of $Z$ are
log-transformed word counts for the problem and solution, the
log-transformed number of steps, six subject indicators, problem difficulty,
and recorded collection generation. The test asks whether the probability
of an annotator finding an error depends on the joint representation of the problem
and solution beyond these features.

CLEAR-CF gives an aggregated $p$-value of $0.0681$, close to the
$0.05$ significance threshold and smaller than those of CLEAR-SS
($0.2921$), PCM ($0.2911$), and $\mathrm{PCM}_M$ ($0.2359$)
(Table~\ref{tab:application-aggregate}). Across the
200 runs, CLEAR-CF gives $p<0.05$ in 117 runs, compared with
63 for CLEAR-SS, 65 for PCM, and 30 for $\mathrm{PCM}_M$
(Table~\ref*{tab:application-splits}). 

\subsection{LLM-judge preferences in Chatbot Arena}

We now turn from human feedback to an LLM judge's preferences.
Chatbot Arena records comparisons between two responses to the same prompt
\citep{chiang2024chatbot,zheng2023judging}. Using response pairs and
LLM judgments supplied by \citet{zhou2026humananchored}, we analyze
$N=3,000$ pairs of responses to distinct initial prompts.
We use Mistral-7B-Instruct-v0.1 as the LLM judge.
The sample construction is detailed in Supplementary
Material~\ref*{app:application-details}.
The outcome $Y$ records the Mistral judge's preference. In the supplied A/B order, $Y=1$ means
the judge prefers A, while $Y=0$ means it prefers B.

The predictor $X$ combines the embeddings of responses A and B into
a normalized pair representation. The 25
coordinates of $Z$ are log-transformed word and character counts for the
prompt and both responses, together with 19 signed contrasts for their
generating models. Each contrast is the model's indicator for response A
minus its indicator for response B. We test whether the representation of the response pair helps
predict this judge's preference after adjustment for lengths and the
models that produced the answers.

After adjustment for lengths and generating model identities, both CLEAR
procedures detect an association between Mistral's preference and the
response pair representation. CLEAR-SS and CLEAR-CF give aggregated
$p$-values of
$0.0256$ and $0.0008$, compared with $0.2093$ for PCM and $0.1892$ for
$\mathrm{PCM}_M$ (Table~\ref{tab:application-aggregate}).
CLEAR-CF gives $p<0.05$ in all 200 runs,
compared with 61 for PCM and 37 for $\mathrm{PCM}_M$
(Table~\ref*{tab:application-splits}).

\begin{center}
\begin{minipage}{\textwidth}
\singlespacing\centering
\captionsetup{hypcap=false}
\captionof{table}{Aggregated nominal $p$-values on each fixed application sample,
using 200 runs for each procedure, with
$\min\{1,2\,\operatorname{median}(p_1,\ldots,p_R)\}$.
Bold values are below $0.05$.}
\label{tab:application-aggregate}
\begin{tabular}{lr rrrr}
\toprule
Application & $N$ & CLEAR-SS & CLEAR-CF & PCM & $\mathrm{PCM}_M$ \\
\midrule
OpenAssistant & 3,400 & \textbf{0.0356} & \textbf{0.0020} & 0.0771 & 0.0934 \\
\addlinespace[0.5em]
PRM800K & 1,000 & 0.2921 & 0.0681 & 0.2911 & 0.2359 \\
\addlinespace[0.5em]
Arena: Mistral & 3,000 & \textbf{0.0256} & \textbf{0.0008} & 0.2093 & 0.1892 \\
\bottomrule
\end{tabular}
\end{minipage}
\end{center}

These findings also have implications for downstream analyses of LLM feedback. In OpenAssistant, the association between response embeddings and human quality ratings suggests that response content contains information relevant to reward modeling beyond length and annotation count. In Arena, Mistral's preferences remain associated with response embeddings after adjusting for length and generating model identities, indicating that length and aggregate preference rates provide only a partial picture of the judge's behavior. Examining the response characteristics underlying these associations may help clarify what human feedback and LLM judges reward, and inform subsequent analyses of reward models and model comparisons.

\section{Discussion}
\label{sec:discussion}

The power of CLEAR depends on how the conditional mean signal aligns
with the graph on $X$. The power results in
Section~\ref{subsec:power-theory} express this requirement
through the graph Rayleigh quotient. When the residual mean changes
sign with $Z$, observations close in $X$ can have opposite signals.
Their negative edge products can offset the positive products from
pairs with matching signs. Some $X$-$Z$ interaction alternatives can
therefore be difficult to detect even when the conditional mean depends
on $X$.

Constructing graphs on $(X,Z)$ may help detect such alternatives by
favoring pairs that are close in both variables. For a signal that
varies smoothly in $(X,Z)$, this could preserve local agreement in
residual means even when its direction changes across values of $Z$.
The baseline regression would still be fitted on $Z$, while the joint
graph would determine which residual products enter the statistic.
A practical question is how to balance the $X$ and $Z$ components of
the dissimilarity, particularly when they have different dimensions or
scales.

Another extension is to use more folds and construct a single graph on all
observations. This would provide more training data for each baseline fit
and allow observations in different folds to be neighbors. Establishing a
Gaussian null limit for this construction would require accounting for the
dependence between fitted residuals along edges connecting different folds.

\subsubsection*{Funding}
This work received no specific funding.

\subsubsection*{Disclosure statement}
The authors report no competing interests.

\subsubsection*{Declaration of generative AI use}
The authors used ChatGPT (GPT-6) and OpenAI Codex (GPT-6.1 Sol) for language
editing, coding assistance, and checking mathematical arguments. The authors verified the originality
and accuracy of the content, checked the tools' terms of use, and take full
responsibility for the manuscript and references.

\nocite{bian2026partial,cai2025test,cai2024mates,cai2022model,chatterjee2024kernel,chatterjee2026conditional,chen2018weighted,chen2017new,chen2024odin,chiang2024chatbot,chu2019asymptotic,dai2022significance,delgado2001significance,dubois2024length,fan1996consistent,friedman1977algorithm,friedman1979multivariate,Hall1980MartingaleLT,hollmann2025tabpfn,kaddour2021causal,Kang2016ScalaronImageRV,kopf2023openassistant,lavergne2015significance,lavergne2000nonparametric,lee2024frameworks,lightman2024verify,liu2026graphbmissing,liu2024generalized,lundborg2024projected,meinshausen2009pvalues,Morris2022SCALARON,ouyang2022training,reimers2019sentence,rosenbaum2005exact,schilling1986multivariate,shuman2013emerging,vonluxburg2014hitting,williamson2023general,Yan2025MachineLearningAssistedCO,zhang2025testing,zheng2023judging,zhou2023ranking,zhou2024changepoint,zhou2026humananchored,zhu2025wise}
\begingroup
\setstretch{1.2}
\setlength{\bibsep}{6pt}

\bibliographystyle{chicago}
\bibliography{Bibliography}

\endgroup

\end{document}